\documentclass[10pt,conference]{IEEEtran}
\IEEEoverridecommandlockouts
\usepackage{cite}
\usepackage{amsmath,amssymb,amsfonts}
\usepackage{algorithmic}
\usepackage{graphicx}
\usepackage{tikz}
\usepackage{textcomp}
\usepackage{xcolor}
\usepackage{soul}

\usepackage{listings}
\usepackage{color}

\definecolor{dkgreen}{rgb}{0,0.6,0}
\definecolor{gray}{rgb}{0.5,0.5,0.5}
\definecolor{mauve}{rgb}{0.58,0,0.82}

\def\BibTeX{{\rm B\kern-.05em{\sci\kern-.025em b}\kern-.08em
    T\kern-.1667em\lower.7ex\hbox{E}\kern-.125emX}}

\newcommand{\tinysection}[1]{\noindent \textbf{#1.}~}

\begin{document}

\title{Automated User Experience Testing through Multi-Dimensional Performance Impact Analysis}

\author{\IEEEauthorblockN{Chidera Biringa, G\"{o}khan Kul}
\IEEEauthorblockA{
\textit{University of Massachusetts Dartmouth}\\
Dartmouth, MA, United States \\
\{\textit{cbiringa}, \textit{gkul}\}@umassd.edu}
}

\maketitle

\begin{abstract}
Although there are many automated software testing suites, they usually focus on unit, system, and interface testing. However, especially software updates such as new security features have the potential to diminish user experience. In this paper, we propose a novel automated user experience testing methodology that learns how code changes impact the time unit and system tests take, and extrapolate user experience changes based on this information. Such a tool can be integrated into existing continuous integration pipelines, and it provides software teams immediate user experience feedback. We construct a feature set from lexical, layout, and syntactic characteristics of the code, and using Abstract Syntax Tree-Based Embeddings, we can calculate the approximate semantic distance to feed into a machine learning algorithm. In our experiments, we use several regression methods to estimate the time impact of software updates. Our open-source tool achieved 3.7\% mean absolute error rate with a random forest regressor.
\end{abstract}


\section{Introduction}
Throughout the Software Development Life Cycle (SDLC), requirements are frequently modified, and modifications are often required after deployment~\cite{Jeong2018}. Some reasons for these modifications include optimization requirements, ever-changing client and user needs, security updates, bug fixes, and feedback from customers based on user experience.

To keep up with the development, software development teams usually establish an extensive continuous integration (CI) and continuous delivery pipeline to automatically test and verify that the software still functions as expected after developer updates. Some tests, such as user experience regression testing, may not be easily automatized. Consequently, it strains software test teams when the software changes rapidly.

When overlooked, user experience in certain functions may diminish due to degrading performance, especially if the user experience testers do not keep track of the changes made to this functionality. This problem becomes evident when the updates made are in specific areas, such as software security which typically requires supplementary computing resources due to key establishment, encryption, decryption, and authorization certificates. An effective testing mechanism must account for the impact of the change in software and the increase in software complexity. 
Otherwise, based on their previous experiences, users stop installing software updates, which will effectively create a large attack surface on the user's device~\cite{Nikitin2017}.

In this paper, we propose a novel automated user experience testing methodology as a part of the CI pipeline.  Our proposed model learns the impact of changes made on the software code and estimates the performance impact when developers attempt to update the codebase. Hence, a software development team can make decisions about adding new features to the software by considering an increase in running time leading to a diminished user experience, which provides them with concrete information about the time-cost of features to the user experience.

To achieve this goal, we analyzed the impact of several versions of software updates in different SDLC phases to provide insights and decision support for software development teams. We created a novel feature set that represents the programming code stylometric characteristics of the Java programming language. We also created an Abstract Syntax Tree-Based Embedding (ASTBE) encoding structure to leverage semantic distance provided by the embedding models and to use the combination of mapped vector features and manually generated features as primary inputs in our machine learning model. Our ASTBE and Random Forest methodology were able to infer the performance impact of security updates in software, with an average error rate of 3.7\% on our dataset, collected from public Java repositories on GitHub.

Concretely, the contributions of this paper are as follows. (1) We propose a novel automated user experience testing method, and (2) We identify challenges to provide ideal conditions for the testing method.

\section{Methodology}
\label{sec:method}
The primary goal of our research is to develop a system capable of inferring the performance of pending updates in software as a way of automated user experience testing. Machine learning (ML) models are apparent techniques used to approach such problems. However, the performance of ML-aided systems is heavily dependent on the quality of the data and how that data is processed for ML algorithmic experimentation.

\begin{figure}[ht]
\centering
\includegraphics[width=\columnwidth]{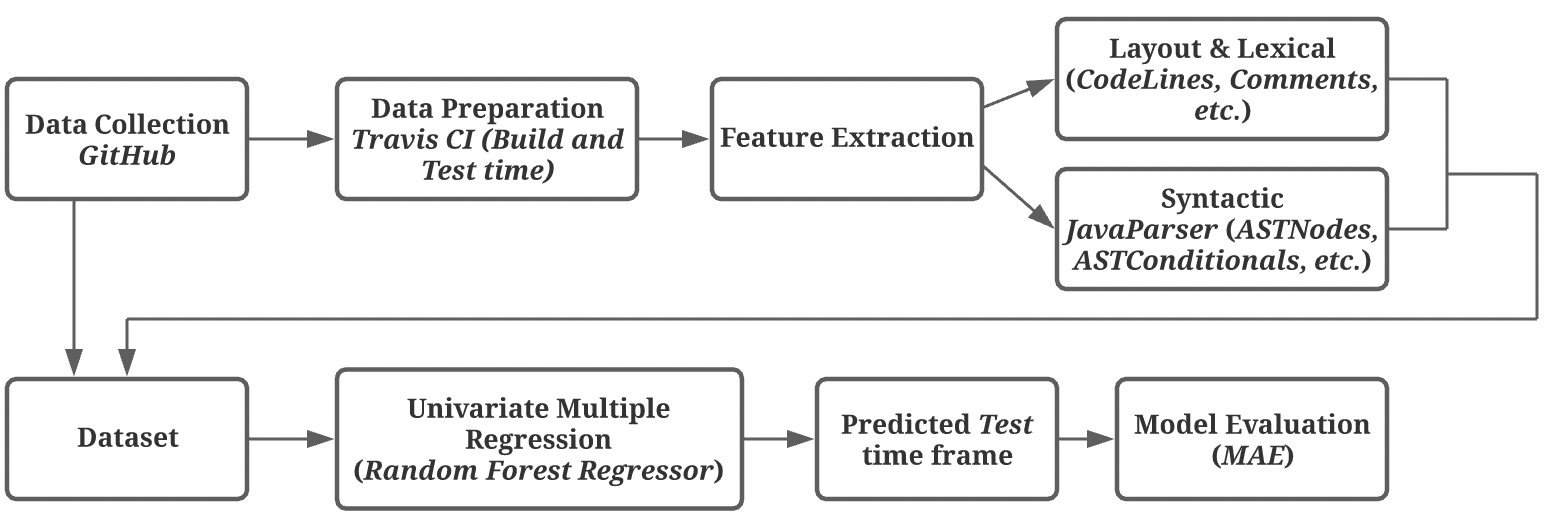}
\caption{Overview of methodology}
\label{fig:overview}
\vspace{-4mm}
\end{figure}

Figure~\ref{fig:overview} lays out the overall methodology of our system.

We assume that the software being developed (repository) is already deployed to a CI pipeline. To automatically calculate the time it takes to run the functional test cases available, we integrate a timer on each test case.
The measured times from the test cases establish our baseline. The source code in the repository is analyzed using code stylometry methods to extract programming language-dependent features. Extracted features and the results of the build and test timeframe of repositories are used to create a final dataset, containing extracted features (Section~\ref{sec:features}). 

The data is used as input in an ML model for predictions -- in our case, a random forest regressor algorithm (Section~\ref{sec:random}). Although we explain our methodology using random forest regressor for conciseness, in Section~\ref{sec:experiments}, we compare several other alternatives.

The code for our methodology is available for anyone to reproduce our results~\footnote{https://github.com/PADLab/MPSS\_Basic}.


\subsection{Features}
\label{sec:features}
Feature engineering is an important step in building the predictive components of intelligent systems~\cite{3}. The performance of our system is heavily dependent on the influence of extracted features. Hence, it is imperative to attentively select and augment features that accurately represent our problem~\cite{4}. Performance analysis of software updates in SDLC is unattainable without the careful representation of source code that denotes the programming preference of contributors to a repository. This problem can be solved by adopting stylometry analysis techniques and recommendations. 

Following Caliskan et al.'s~\cite{caliskan2015anonymizing} and Abbasi et al.'s~\cite{5} research, we extracted code features by applying statistical methods, regular expression, and a tool used to parse Java code. We combined \emph{lexical, layout,} and \emph{syntactic} features to obtain a robust stylometry feature set. We also created a unique \emph{Abstract Syntax Tree-Based Embedding (ASTBE)} that will be the primary input data in creating a deep neural network model in our future work.


\subsubsection{Layout and Lexical Features}
Layout features represent the organizational structure of the code such as an author's preference for space over tabs. Lexical features are the semantic representation of the programming language; they represent the context and preferential formal adoptions of code syntax. We began by applying statistical analysis and regular expression to extract \emph{Lexical} and \emph{Layout} features. A \texttt{Regular Expression} is the concatenation of characters that represents a regular language search pattern~\cite{li2008regular}. We wrote patterns that matched the features. For example, consider the regular expression pattern shown in Figure 2. The search pattern contains a collection of characters that matches \emph{Java Comments} in a source code. After the successful pattern matching of a feature, we extracted the matched features by applying the equation shown in \ref{fe_eqn}.

\begin{figure}[ht]
\begin{lstlisting}
"/\*(.|[\r\n])*?\*/|//.*"
\end{lstlisting}
\vspace{-4mm}
\caption{Regular Expression Listing}
\vspace{-4mm}
\end{figure}

\begin{equation}
\label{fe_eqn}
Feature Extractor(FE) = \log_{10}\left(\frac{\sum\limits_{i=0}^{n}a_i}{\vert x \vert}\right) 
\end{equation}

$\sum\limits_{i=0}^{n}a_i$ represents the summation of feature occurrences in a single file. For example, we parse a single file and generate an AST. Next, \emph{loop statements} features are extracted by counting the number of times a \emph{For, ForEach and While loops} occur. We divided the feature by $\vert x \vert$ (character length of a file). Finally, we reduce the skew of feature distribution by taking the log of the result.  

\begin{table}[ht]
\begin{minipage}{\columnwidth}
\centering
\begin{tabular}{ll} 
\hline
\textbf{Features} & \textbf{Count} \\
\hline 
Imports & 1 \\
\hline
Comments & 3 \\
\hline
Keywords & 2 \\ 
\hline
Methods & 3 \\
\hline 
Unigrams & *ngram dependent \\ 
\hline 
\end{tabular}
\vspace{0.1cm}
\caption{Lexical Features} 
\end{minipage}
\hspace{-3cm}
\begin{minipage}{\columnwidth}
\centering
\begin{tabular}{ll}
\hline 
\textbf{Features} & \textbf{Count} \\  
\hline 
CodeLines(AVG) & 2 \\
\hline
CodeLines(SD) & 1 \\
\hline
EmptyLines & 2 \\
\hline
WhiteSpace & 1 \\
\hline
Tabs & 0 \\ 
\hline 
\end{tabular}
\vspace{0.1cm}
\caption{Layout Features}
\end{minipage} 
\end{table}

\subsubsection{Syntactic Features}
Syntactic features are programming language dependent and provide context on a programmer's style-based preferential characteristics present in a source code file. Syntactic features are Context-Free Language-specific and are derived using Context-Free Grammar(CFG). CFG expresses specific characteristics of a source code, For example, the recursive structure of a code snippet. A CFG is a principle of \emph{formal language}, it denotes the collection of possible languages given a grammar. A CFG comprises of 4 tuples: $\mathnormal{(V, \Sigma, R, S)}$~\cite{sipser1996introduction}. $\mathnormal{V}$ is a finite set of \textit{variables}. $\Sigma$ is a finite set of terminals disconnected from $\mathnormal{V}$. $\mathnormal{R}$ is a finite set of production rules. $\mathnormal{A \rightarrow X}$ where $\mathnormal{A \in V}$, $\mathnormal{X  \text{ } \in  \text{ } (V  \text{ } \cup \text{ } \Sigma)}$ and $\mathnormal{ S  \in  V}$ is the beginning variable.

Extracting syntactic features is a decidable task contingent on the generation and analysis of a parse tree. Hence, we used a parser capable of generating an Abstract Syntax Tree (AST). An AST is a tree data structure of a programming language~\cite{neamtiu2005understanding} and is the highest syntactic component of a single source code file. We used \emph{JavaParser}~\cite{smith2017javaparser} to parse source code files and analyze its ASTs. The \emph{Count} column in Table III represents the influence of a feature in the source code.

\begin{figure}[ht]
\centering
\begin{lstlisting}
public boolean checkPositive(int integer) {
if (integer > 0) return true;
else return false;
}
\end{lstlisting}

\big\downarrow
\vspace{6mm}

\tikzset { level 2/.style={sibling distance=5.5cm}, level 6/.style={sibling distance=5.7cm}}
\resizebox{\linewidth}{!}{%
\begin{tikzpicture}[->, every node/.style={rectangle, draw=black}, level/.style={sibling distance = 2cm, level distance = 0.9cm}]
\small 
\node{root(CompilationUnit)}
		child{ node{MethodDeclaration}
			child{ node{VariableDeclaration(kind='int')} }
			child{ node{checkPositive(VariableDeclarator)} 
				child{ node{MethodExpression}
					child{ node{BlockStatement} 
						child{ node{IfStatement}
							child{ node{BinaryExpression(Operator='$>$')} 
								child{ node{Identifier(name='integer')}}
								child[level distance=1.7cm, sibling distance = 6.2cm]{ node{Literal(value='integer')}}
							}
							child{ node{ReturnStatement} 
								child{ node{ Literal(value='true')} }
							}
							child{ node{ReturnStatement} 
								child{ node{Literal(value='false')} }
							}
						}
					}
				}
			}
		}
	;
\end{tikzpicture}
}
\caption{Sample code listing and its equivalent Abstract Syntax Tree (AST)}
\end{figure}
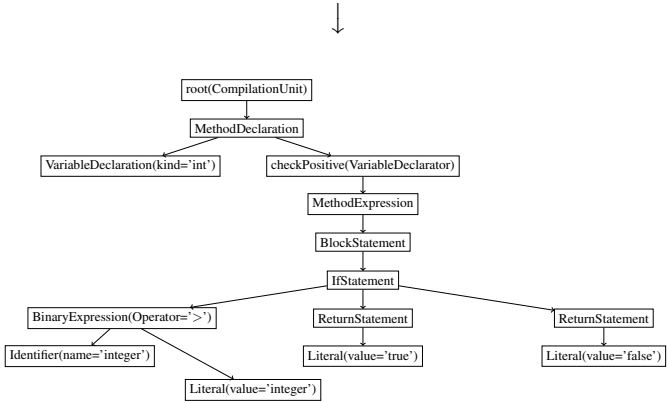
                                    
Figure 3 displays an example of a simple code snippet to check whether an integer is positive or negative with its corresponding AST. The method contains language-dependent variable declarations, expressions, and statements. Individual code lines and language constructs are equivalent node representations decoded using an AST. The analysis of ASTs can obtain characteristics common in programming habits and preferences from source code.

\begin{table}[ht]
\caption{Syntactic Features} 
\centering
\begin{tabular}{ll} 
\hline
\textbf{Features} & \textbf{Count} \\ 
\hline
ASTConditionals & 3 \\ 
\hline
ASTLiterals & 3 \\
\hline
ASTLoops & 3 \\
\hline
ASTNodes & 1 \\
\hline
ASTTFIDF & *dynamic \\
\hline
\end{tabular}
\vspace{0.1cm}
\end{table}

\vspace{-6mm}

\subsection{Abstract Syntax Tree Based Embedding(ASTBE)}
\label{sec:random}
We created an ASTBE encoding structure that works for ASTs. This architecture modifies the Continuous Bag of words(CBOW) model to contain specific node types derived from processing our ASTs. There are three reasons why we created an ASTBE. First, to leverage the semantic distance provided by embedding models, preserving the relationship between nodes in vector space. Second, to leverage the transferability and extraction of Deep learning features, mapped vector features and manually generated features served as independent or co-dependent primary inputs for our model. Third, to ascertain the applicability of ASTBE in the problem domain of impact analysis of code updates on software. We used the \emph{JavaParser} library to parse and generate ASTs. Next, we traversed the trees using a Preorder traversal algorithm and selected representative nodes. We adopted node selection and encoding techniques designed by Zhang \textit{et al.}~\cite{zhang2019novel}. Finally, we converted selected nodes to vectors.

\tinysection{Random Forest Regressor} There are many regression models. Random Forest Regressor is an ensemble learning method built from a collection of decision trees, also called an estimator~\cite{10}. These estimators were built using randomly sampled features of data. We used 100 decision trees for our forest and handled over-fitting by setting the tree depth not to grow beyond two levels. It ensures generalization and maintains the performance stability of the test set.

\section{Evaluation}
\label{sec:experiments}
\emph{Travis CI} is an open-source continuous integration service used to build and test software hosted on GitHub and other source code management and distributed version control platforms~\cite{beller2017oops}.
We linked our repositories to \emph{Travis CI}.
We created and deployed build and test scripts into our CI pipeline using a \emph{.travis.yml} file. The file was tasked with compiling and executing unit test cases present in each repository. We started with a single repository containing five \texttt{git} branches. Individual branches of the repository contained a total of 42 Java files. The files were concatenated with the results of repository build and test time derived from our CI pipeline to create our full primary dataset. We left out files in the repository containing text other than Java code. We recognize that there might exist a significant code homogeneity between branches. However, our focus is on the increase or decrease in repository performance time.

\tinysection{Data collection} The primary source of raw data used in this research comes from public Java repositories on GitHub. We applied four filters to in the identification and selection of candidate repositories. First, the repository must be written using the Java programming language. Second, the repository must contain complete code files, i.e., contain compilable code. Third, the repository must have existing unit and system test cases. Fourth, the repository must contain a considerable number of commits, stars, and contributors. These selected thresholds ensure that we achieve a representative collection of candidate repositories used in creating a dataset. Selected repositories were forked and cloned. Given the commits hashes, we used \texttt{git reset} and \texttt{git log} commands to browse and restore repository histories at update checkpoints. Finally, \texttt{git} branches were created using the above-stated checkpoints and deployed to a CI pipeline to be built and tested. We share the repository we used in our evaluation for reproducibility purposes~\footnote{ https://github.com/se-edu/addressbook-level2}, although our evaluation can be performed on any repository that satisfies the selection criteria.

\begin{figure}[ht]
\centering
\includegraphics[width=\linewidth]{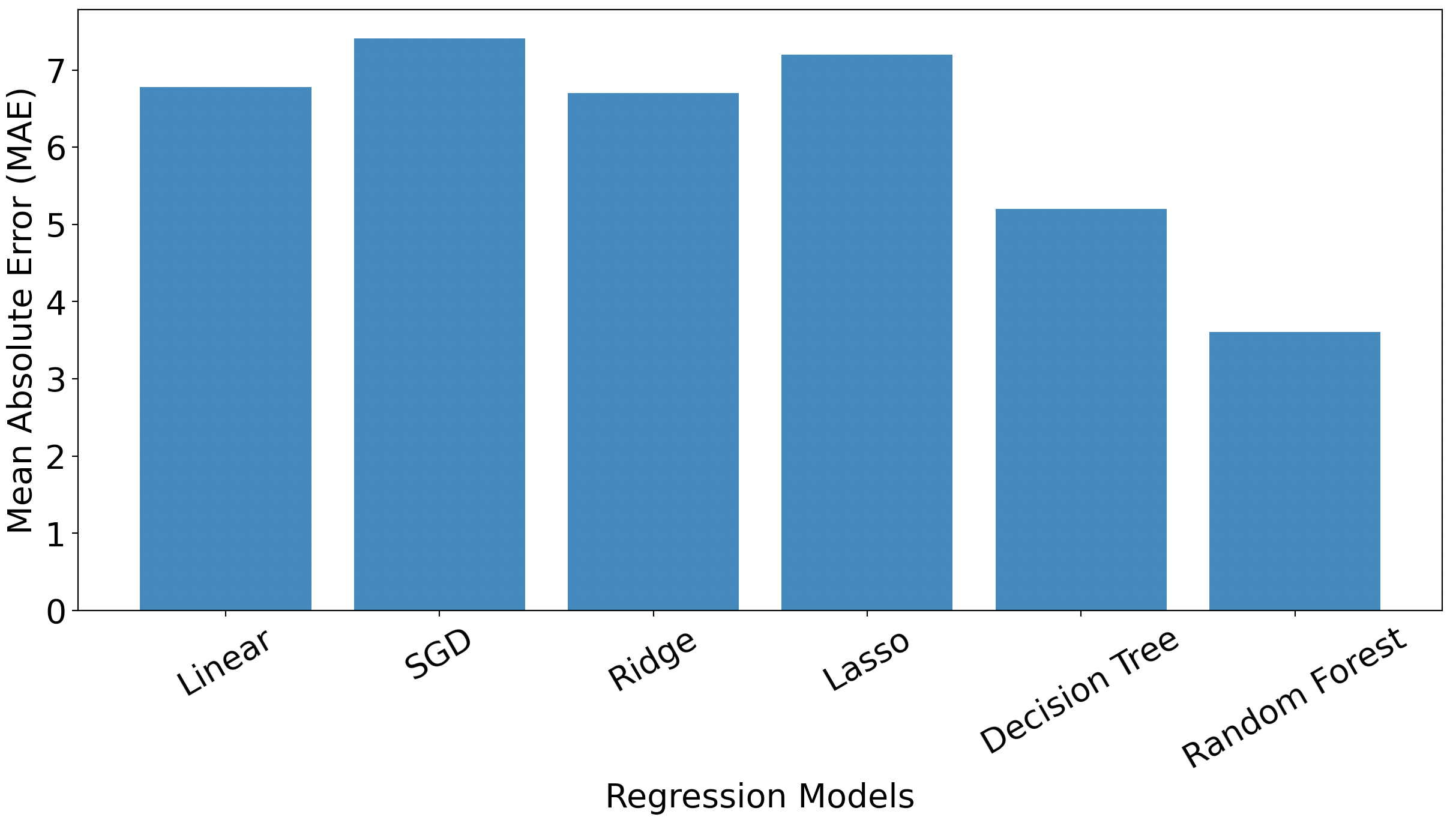}
\vspace{-5mm}
\caption{MAE of regression models (lower is better)}
\label{fig:MAE}
\end{figure}


\tinysection{Data preparation} We used CI pipeline as a data preparation tool to automatically build and test multiple branches of our repositories. CI is an agile software development tool used in rapidly integrating programming code into distributed and shared repositories~\cite{2}. This integration facilitates the automatic building and testing of code updates made to the software. We extracted time data in different states of the repository, such as the build and test time of current code states (CCS). 

\tinysection{Experimentation Setup} We created a pipeline of various regression models. Extracted features and timeframe data from the CI pipeline are primary inputs in the models. Our dataset comprises of 13 predictor variables, \emph{Test(sec)} target variable, and 210 observations. We used a \emph{k-fold cross-validation} resampling method~\cite{browne2000cross} and split the data into training and testing set representing equal folds. The splits are stratified by the testing sets(\emph{Test time feature}). Hence, the size of k is dependent on the dataset. With a different fold, the data is cross-validated for \emph{10 iterations}. The average cross-validated error is applied to estimate out-of-sample mean absolute error.

\tinysection{Evaluation} We use Mean Absolute Error (MAE), given in Equation~\ref{eq:MAE}, to test the performance of a given regression model. It calculates the mean size of errors in a collection of predictions. MAE does not consider the directional relationship between predicted and actual observations. It is only concern with the absolute difference. Hence, individual values have equivalent significance~\cite{chai2014root}. MAE has the advantage of punishing smaller errors, which is particularly significant in the problem domain, because we are predicting time intervals with minuscule differences. We used MAE to evaluate the performance of our cross-validated predictions. 

\begin{equation}
\label{eq:MAE}
Mean Absolute Error (MAE) = (\frac{1}{n})\sum_{i=1}^{n}\left | y_{i} - x_{i} \right |
\end{equation}

We have tested our pipeline using several regression models. The Linear Regression model achieved 6.79\%, SGD Regressor achieved 7.47\%, Ridge Regression achieved 6.71\%, Lasso Regression achieved 7.20\%, Decision Tree Regressor achieved 5.19\%, Random Forest Regressor achieved 3.70\% error rate on the average cross-validated predictions, as shown in Figure~\ref{fig:MAE}.


\section{Conclusion and Future Work}
\label{sec:conclusion}
In this paper, we proposed a novel automated user experience testing tool that can integrate into existing continuous integration pipelines to provide immediate feedback to software development teams. We tested this method on an existing GitHub repository and achieved a 3.7\% MAE rate.

While our results are promising, our scope for this paper is limited, and we are planning to expand our work in several directions. First, we acknowledge that the codebase needs to have a significant and carefully placed unit, system, and integration tests for our system to collect timing data from. We will need to evaluate when performing our estimations is infeasible and what conditions create the ideal environment. Second, we plan to expand our raw data by including a wide variety of repositories from GitHub. Third, we will expand our work to intelligently estimate its accuracy by applying our model with each commit and learning from the change in error rate. Lastly, we will explore deep neural network models that have the potential to yield higher accuracy.

\section*{Acknowledgment}
This work has been funded by UMass Dartmouth Cybersecurity Center. Usual disclaimers apply.

\bibliographystyle{IEEEtran}
\bibliography{references}

\end{document}